\documentstyle[12pt]{article}
\def\qed{\qquad\vrule height4pt width6pt depth2pt}

\def\1{{\chi}}

\begin{document}
\title {{Quantum Observable Generalized Orthoalgebras}\thanks{This project is supported by Natural Science
Found of China (11571307, 11101108 and 11171301) and by the Doctoral Programs Foundation of the Ministry of
Education of China (20120101110050).
}}
\author {Qiang Lei $^{1}$, Weihua Liu $^{2}$, Zhe Liu $^{3\dag}$, Junde Wu $^{4}$\date{}\thanks{Corresponding author: liu.z@duke.edu; liuweih@indiana.edu; wjd@zju.edu.cn}}
\maketitle

\noindent$^1${\small\it Department of Mathematics, Harbin Institute of Technology, Harbin 150001, China}

\noindent$^2${\small\it Department of Mathematics,  Indianan University, Bloomington, Indiana 47401, USA}

\noindent$^3${\small\it Division of Natural Sciences, Duke Kunshan University, Kunshan 215316, China}

\noindent$^4${\small\it School of Mathematical Science, Zhejiang University, Hangzhou 310027, China}

\vskip0.2in

 {\noindent {\bf Abstract}. \small\it {Let ${\cal
S}(\mathcal{H})$ denote the set of all self-adjoint operators (not necessarily bounded) on a Hilbert space
$\mathcal{H}$, which is the set of all physical quantities on a quantum system $\mathcal{H}$.
We introduce a binary relation $\bot$ on ${\cal
S}(\mathcal{H})$. We  show that if $A\bot B$, then $A$ and $B$ are affiliated with some
abelian von Neumann algebra. The relation $\bot$
induces a partial algebraic operation $\oplus$ on ${\cal
S}(\mathcal{H})$. We prove that $({\cal S}({\mathcal{H}}), \bot, \oplus, 0)$ is a generalized
orthoalgebra.  This algebra is a generalization of the famous Birkhoff\,--\,von Neumann quantum logic model. 
It establishes a mathematical structure on all physical quantities on $\mathcal{H}$. In particular, we note that $({\cal S}({\mathcal{H}}), \bot, \oplus, 0)$ has a
partial order $\preceq$, and prove that $A\preceq B$ if and only if
$A$ has a value in $\Delta$ implies that $B$ has a value in $\Delta$
for every Borel set $\Delta$ not containing $0$. Moreover, the existence of the infimum $A\wedge B$ and supremum
$A\vee B$ for $A,B\in \mathcal{S}(\mathcal{H})$ (with respect to $\preceq$) is studied, and it is shown
at the end that the position operator $Q$ and momentum operator $P$ in the Heisenberg commutation
relation
satisfy $Q\wedge P=0$.}}

\vskip0.2in

{\noindent{\bf Key Words.} \small\it{Quantum observable, Self-adjoint operator, Generalized orthoalgebra, Order.}}

\vskip0.2in

\section{Introduction}

The sixth problem of Hilbert's Mathematical Problems (Hilbert outlined 23 major mathematical problems
in 1900 to be studied in the coming century) is about
mathematical treatment of the axioms of physics -- ``{\it The investigations on the foundations of
geometry suggest the problem: To treat in the
same manner, by means of axioms, those physical
sciences in which already today mathematics plays
an important part; in the first rank are the theory
of probabilities and mechanics.\/}" In 1933, Kolmogorov axiomatized modern
probability theory (\cite{KOL1}). In Kolmogorov's theory, the set $\cal L$
of experimentally verifiable events form a Boolean
$\sigma$-algebra; therefore, Boolean algebra theory can be used to describe classical logic. However, Kolmogorov's theory does not describe
situations that arise from quantum mechanics, e.g.  the
Heisenberg uncertainty principle (\cite{HEI1}). One of the most
fundamental problems in quantum theory is to find a mathematical description
for the structure of random events in a quantum system. This
was originally studied in 1930s by Birkhoff and von Neumann
(\cite{BN1}). In  von Neumann's approach, a quantum
system is represented by a separable complex Hilbert space
$\mathcal{H}$; each physical quantity is represented by a
self-adjoint operator on $\mathcal{H}$, and is called a quantum
observable. The set of all quantum observables is denoted by
${\cal S}(\mathcal{H})$. Since the spectrum $\sigma(P)$ of a projection
operator $P$ is contained in $\{0, 1\}$, if the truth values (true and false)
for two-valued propositions about the quantum system are
encoded by $0$ and $1$, then these propositions can be represented by
projections on $\mathcal{H}$.  Birkhoff and von Neumann
considered the set $\mathcal{P}(\mathcal{H})$ of all projections on $\mathcal{H}$ as the logic of the quantum system
(\cite{BN1}). If $A\in {\cal S}(\mathcal{H})$ and $P^A$ is the
spectrum measure of $A$,  then for each real Borel set $\Delta$,
$P^A(\Delta)$ represents the event that the values of physical
quantity $A$  are contained in $\Delta$.

Let $\cal L$ be a lattice with two binary operations the supremum
$\vee$ and the infimum $\wedge$. If there are elements $0$ and $I$
in $\cal L$ and a unary operation $': {\cal L}\rightarrow {\cal L}$
such that  $x''=x, x\vee
x'=I, x\wedge x'=0$, and $0\leq x\leq I$ for each $x\in \cal L$, then $({\cal L}, \vee, \wedge, \/', 0, I)$ is
said to be an {\it ortholattice}, and $'$ is said to be an
orthocomplementation operation.

We say that an ortholattice $({\cal L}, \vee, \wedge, \/', 0, I)$
satisfies the orthomodular law if for $x, y\in {\cal L}$,
$$x\leq y\Rightarrow y=x\vee(y\wedge x').$$

\noindent The name of orthomodular was suggested by Kaplansky. An
ortholattice satisfying the orthomodular law is said to be an
{\it orthomodular lattice} (\cite{KA}).
We say an ortholattice $({\cal L}, \vee, \wedge, \/', 0, I)$
satisfies the modular law if for $x, y,z\in {\cal L}$, $$x\leq
y\Rightarrow y\wedge (x\vee z)=x\vee(z\wedge y).$$
An ortholattice satisfying the modular law is said to be a {\it modular
lattice} (\cite{KA}).

Let $p, q\in \mathcal{P}(\mathcal{H})$. We say that $p\leq q$ if $\langle px,x\rangle\leq \langle qx,x\rangle$ for all $x\in \mathcal{H}$.  Then $(\mathcal{P}(\mathcal{H}), \leq)$ is a lattice with respect to the partial order
``$\leq$ " with the minimal element $0$ and the maximal element $I$. If we define  $p'=I-p$, then
$({\cal P}({\cal H}), \vee, \wedge, \/', 0, I)$ is an ortholattice.
Husimi (\cite{HU1}) showed that $({\cal P}({\cal H}), \vee, \wedge, \/', 0, I)$ is an orthomodular lattice.
Also see \cite{KA} for the use of orthomodular lattices in quantum logic.
Birkhoff and von Neumann (\cite{BN1}) showed that if $\mathcal{H}$ is finite dimensional, then
$({\cal P}({\cal H}), \vee, \wedge, \/', 0, I)$ is a modular lattice .

There are properties that clearly distinguish quantum logic from classical logic. Note that each Boolean algebra $\cal A$ (for classical logic) is a distributive ortholattice, that is, for  $x, y, z\in\cal A$, $$x\wedge(y\vee z)=(x\wedge y)\vee (x\wedge z)\quad \mbox{and}\quad x\vee(y\wedge z)=(x\vee y)\wedge (x\vee z),$$
while the distributive law does not hold in $({\cal P}({\cal H}), \vee, \wedge, \/', 0, I)$ (for quantum logic).

Let $({\cal L}, \vee, \wedge, \/', 0, I)$ be an orthomodular lattice.
 We say that $x$ and $y$ satisfy the  binary relation $\bot$ if  $x\leq y'$.  We define a partial operation $\oplus$ on $\cal L$ by
$x\oplus y=x\vee y$ if
$x\bot y$. Then,  we obtain a new algebraic structure $({\cal L},
\bot, \oplus, 0, I)$ with  the following properties:

(OA1) If $x\bot y$, then $y\bot x$ and $x\oplus y=y\oplus x$.

(OA2) If $y\bot z$ and $x\bot (y\oplus z)$, then $x\bot y$,
$(x\oplus y)\bot z$ and $(x\oplus y)\oplus z=x\oplus (y\oplus z)$.

(OA3) For each $x\in \cal L$, there exists a unique $y\in\cal L$
such that $x\bot y$ and $x\oplus y=I$.

(OA4) If $x\bot x$, then $x=0$.

\noindent Foulis, Greechie and R\"{u}ttimann called this structure
$({\cal L}, \bot, \oplus, 0, I)$ an {\it orthoalgebra} (\cite{FGR}).
Kalmbach, Rie\v{c}anov\'{a}, Hedl\'{\i}kov\'{a}, Pulmannov\'{a} and Dvure\v{c}enskij introduced
the following definition ({\cite{KG96,HJ96,DA02}}):

\vskip0.1in

{\bf Definition 1}. A {\it generalized orthoalgebra} $({\cal E}, \bot,
\oplus, 0)$ is a set $\cal E$ with an element $0$, a binary relation
$\bot$, and a partial operation $\oplus$, such that if $x\bot y$,
then $x\oplus y$ is defined and satisfies the following conditions:

(OA1). If $x\bot y$, then $y\bot x$ and $x\oplus y=y\oplus x$.

(OA2). If $y\bot z$ and $x\bot (y\oplus z)$, then $x\bot y$,
$(x\oplus y)\bot z$ and $(x\oplus y)\oplus z=x\oplus (y\oplus z)$.

(OA4). If $x\bot x$, then $x=0$.

(GOA1). If $x\bot y$, $x\bot z$ and $x\oplus y=x\oplus z$, then
$y=z$.

(GOA2). $x\bot 0$ and $x\oplus 0=x$ for all $x\in\cal E$.

(GOA3). If $x\bot y$ and $x\oplus y=0$, then $x=y=0$.

\vskip0.1in

Let $({\cal E}, \bot,
\oplus, 0)$ be a generalized orthoalgebra. For $a, b\in\cal E$, if there is a $c\in\cal E$ such that $a\bot c$
and $a\oplus c=b$, then we say that $a\preceq b$. It can be shown
that $\preceq$ is a partial order. Moreover, $x\bot y$ if and only if $x\leq
y'$ (\cite{DA02}).
Generalized orthoalgebras are very important models of quantum logic (\cite{DA02}).

In \cite{GS06}, Gudder defined a binary relation $\bot$ on the set
$\mathcal{S}_{\bf b}(\mathcal{H})$ of all bounded self-adjoint operators on $\mathcal{H}$ by
$A\bot B$ once $AB=0$, and then define $A\oplus B=A+B$. However, many of the
operators that arise naturally in physics are not bounded. For example, in
Heisenberg's commutation relation, a fundamental relation in quantum mechanics, $QP-PQ=-i\hbar I$, the position operator $Q$ and the momentum operator
$P$ cannot be realized by bounded operators (see \cite{KRV12} for a full account on this). Therefore, it is necessary to study unbounded operators and, in particular, the set ${\cal S}(\mathcal{H})$ of all self-adjoint
(possibly unbounded) operators on $\mathcal{H}$.

In this paper, we introduce a binary relation $\bot$ on ${\cal
S}(\mathcal{H})$. For $A, B \in {\cal
S}(\mathcal{H})$, we say $A\bot B$ if and only if $\overline{ran}(A)$ is orthogonal
to $\overline{ran}(B)$, where $\overline{ran}(\cdot)$ denotes the closure of the range of an operator. If
$A\bot B$, define $A\oplus B = A+B$. We show that if
$A\bot B$, then $A$ and $B$ are affiliated with
some abelian von Neumann algebra. Moreover, we show that $({\cal
S}({\mathcal{H}}), \bot, \oplus, 0)$ is a generalized orthoalgebra. In this way, we establish a new quantum logic structure on all physics quantities of the quantum system $\mathcal{H}$. Note that the generalized orthoalgebra $({\cal
S}({\mathcal{H}}), \bot, \oplus, 0)$ has a nature partial order $\preceq$. We show that
$A\preceq B$ if and only if $A$ has a value in $\Delta$ implies that $B$ has a
value in $\Delta$ for every Borel set $\Delta$ not containing
$0$. The existence of the infimum $A\wedge B$ and
supremum $A\vee B$ for $A,B\in \mathcal{S}(\mathcal{H})$ with respect to
$\preceq$ is also studied. At the end,  we show that the position operator $Q$ and momentum
operator $P$ satisfy $Q\wedge P=0$ with respect to $\preceq$.

\section{Definitions and Facts of Self-adjoint Operators}

We first  recall some elementary concepts and facts of unbounded linear operators (see Section 4 of \cite{KRV12}
for a brief summary, and Sections 2.7 and 5.6 of \cite{KRV97} and Section 6.1
of \cite{KRV86} for more details). A
linear operator $A$ we consider will have a domain
${\mathcal{D}}(A)$ that is dense in $\mathcal{H}$.
Given two linear operators $A: {\mathcal{D}}(A)\rightarrow
\mathcal{H}$ and $B:{\mathcal{D}}(B)\rightarrow \mathcal{H}$, we
write $A\subseteq B$ and say that $B$ is an extension of $A$, if ${\mathcal{D}}(A)\subseteq
{\mathcal{D}}(B)$ and $Ax=Bx$ for all $x\in {\mathcal{D}}(A)$. For a linear operator
$A:{\mathcal{D}}(A)\rightarrow
\mathcal{H}$, the adjoint of $A$, denoted by $A^*$, is defined as follows.
Its domain consists of those vectors $y$ in $\mathcal{H}$ such that
for some $y^*$ in $\mathcal{H}$, $\langle x, y^*\rangle=\langle Ax,y\rangle$
for all $x\in {\mathcal{D}}(A)$, and $A^*y=y^*$ for each $y\in
{\mathcal{D}}(A^*)$. We say that $A$ is symmetric if $A\subseteq A^*$, and
self-adjoint if $A=A^*$. If $A\subseteq B$, then $B^*\subseteq A^*$.
If $A\subseteq B$ with $A$ self-adjoint and $B$ symmetric, then $A=B$
(in this case $A$ has no proper symmetric extension, that is, a self-adjoint operator
is maximal symmetric).
We say that $A$ is closed if its
graph $G(A)=\{(x, Tx)|x\in {\mathcal{D}}(A)\}$ is closed, and $A$ is closable (or preclosed) if there exists a closed linear operator $B$ such that its graph is the closure of the graph of $A$, $\overline{G(A)}=G(B)$. In this case, $B$
is called the closure of $A$, denoted by $\overline{A}$.
Self-adjoint operator are closed and symmetric operators are closable.
If $A$ is closable, then $(\overline{A})^*=A^*$.
If $A$ is closed and $\overline{G(A|_{{\mathcal{D}}_0})}=G(A)$, where ${\mathcal{D}}_0$
is a dense linear subspace of ${\mathcal{D}}(A)$, we say that ${\mathcal{D}}_0$ is a core for $A$.

A family $\{E_{\lambda}\}$ of projections indexed by $\mathbf{R}$, satisfying

(i) $\wedge_{\lambda\in \mathbf{R}}E_{\lambda}=0$ and $\vee_{\lambda\in \mathbf{R}}E_{\lambda}=I$,

(ii) $E_{\lambda_1}\leq E_{\lambda_2}$ if $\lambda_1\leq\lambda_2$,

(iii) $\wedge_{\lambda\geq \lambda_1}E_{\lambda}=E_{\lambda_1}$,

\noindent is said to be a resolution of the identity.
The following is a spectral theorem for self-adjoint operators.

\vskip0.1in

{\bf Lemma 1}. \label{l:204} If $A$ is a self-adjoint
operator on $\mathcal{H}$, then there is a unique (projection-valued)
spectral measure $P^A$ defined on all Borel subsets of $\mathbf{R}$ such that
$$A=\int_{\mathbf{R}} \lambda d P^A(\lambda).$$

\vskip0.1in

\noindent If we denote $E^A_\lambda =P^A((-\infty,\lambda])$, then $\{E^A_\lambda\}$ is a resolution of the identity, and it is said to be the resolution of the identity for $A$. Let $F^A_n=E^A_n-E^A_{-n}$. Then $\bigcup_{n=1}^\infty F^A_n(\mathcal{H})$ is a core for $A$, and for each $x\in F^A_n(\mathcal{H})$ and $n\in \mathbf{N}$, $$Ax=\int_{-n}^{n} \lambda dE^A_\lambda
x$$  in the sense of norm convergence of approximating Riemann sums. In addition, for $m,n\in \mathbf{N}$, $F^A_nF^A_m=F^A_mF^A_n$,
$F^A_nA\subseteq AF^A_n$, and $AF^A_nx\rightarrow Ax$ for each $x\in
{\mathcal{D}}(A)$.

\vskip0.1in

{\bf Lemma 2}. If $A$ is a closed, then the null space of $A$, $null(A)=\{x\in
{\mathcal{D}}(A):Ax=0\}$, is a closed subspace of
$\mathcal{H}$. Moreover, $(ran(A))^\perp=null(A^*)$,
$(ran(A^*))^\perp=null(A)$,
$\overline{ran}(A^*A)=\overline{ran}(A^*)$, $null(A^*A)=null(A)$.

\vskip0.1in

Let $P_A$ and $N_A$ denote the projections whose ranges are
$\overline{ran}(A)$ and $null(A)$, respectively.

\vskip0.1in

{\bf Lemma 3}. \label{l:203} Let $A,B\in
\mathcal{S}(\mathcal{H})$. Then $P^A(\{0\})=N_A$,
$P^A({\mathbf{R}}\setminus\{0\})=P_A$, $P_A+N_A=I$ and $P_A\vee
P_B=I-N_A\wedge N_B$.

\vskip0.1in

{\bf Lemma 4} (\cite{ZGQ90}). \label{l:302} Let $A\in
\mathcal{S}(\mathcal{H})$. If $B$ is a bounded linear operator on $\mathcal{H}$ and
$BA\subseteq AB$, then for a Borel set $\Delta \subseteq \mathbf{R}$,
$P^A(\Delta)B=BP^A(\Delta)$.

\vskip0.1in

{\bf Lemma 5.}\label{l:5} Suppose that $A$ and $B$ are densely defined
on $\mathcal{H}$. Then $A^*+B^*\subseteq(A+B)^*$ if $A+B$ is densely defined,
and $B^*A^*\subseteq(AB)^*$ if $AB$ is densely defined.

\vskip0.1in

{\bf Lemma 6.}\label{l:6} If $A$ and $C$ are densely defined closable operators
on $\mathcal{H}$ and $B$ is a bounded operator on $\mathcal{H}$ such that
$A=BC$, then $A^*=C^*B^*$.

\vskip0.1in

{\bf Lemma 7.}\label{l:301} Let $A,B\in \mathcal{S}(\mathcal{H})$. Then the
following statements are equivalent:

(i) $A\perp B$, that is, $\overline{ran}(A)$ is orthogonal
to $\overline{ran}(B)$.

(ii) $\overline{ran}(A)\subseteq null(B)$.

(iii) $\overline{ran}(B)\subseteq null(A)$.

(iv) $AB\subseteq 0$ and
${\mathcal{D}}(AB)={\mathcal{D}}(B)$.

(v) $BA\subseteq 0$ and
${\mathcal{D}}(BA)={\mathcal{D}}(A)$.

{\bf Proof.} Clearly, $(i)\Leftrightarrow (ii)\Leftrightarrow (iii)$.

$(ii)\Leftrightarrow (v)$: Suppose $BA\subseteq 0$ and
${\mathcal{D}}(BA)={\mathcal{D}}(A)$. For each $x\in
{\mathcal{D}}(A)$, $Ax\in {\mathcal{D}}(B)$ and $BAx=0$. That is
$Ax\in null(B)$. So $ran(A)\subseteq null(B)$. Since $null(B)$ is
closed, $\overline{ran}(A)\subseteq null(B)$. Conversely, suppose that
$\overline{ran}(A)\subseteq null(B)$. Then, for each $x\in
{\mathcal{D}}(A)$, $Ax\in null(B)$, we have $x\in {\mathcal{D}}(BA)$ and
$BAx=0$. Therefore, ${\mathcal{D}}(BA)={\mathcal{D}}(A)$ and $BA\subseteq
0$.

Similarly, $(iii)\Leftrightarrow (iv)$. \qed

\section{The Affiliate Relationship}

We say that a closed operator $T$ is {\it affiliated} with
a von Neumann algebra $\mathcal{R}$ and write $T\eta \mathcal{R}$
when $UT=TU$ for each unitary operator $U$ commuting with
$\mathcal{R}$. (Note that the equality
$UT = TU$ means that ${\mathcal D}(UT)(= {\mathcal D}(T))={\mathcal D}(TU)$ and $UTx = TUx$ for each $x\in {\mathcal D}(T)$
and $U$ maps ${\mathcal D}(T)$ onto itself.)

Murray and von Neumann showed (\cite{MvN}) that the family of operators
affiliated with a factor of type II$_1$ (or, more generally, affiliated with a finite von Neumann
algebra, those in which the identity operator is finite) admits surprising operations of addition
and multiplication that suit the formal algebraic manipulations used by the founders of
quantum mechanics in their mathematical model. See Section 6 of \cite{KRV12} for fundamental properties
of affiliated operators.

\vskip0.1in

{\bf Lemma 8} (\cite{KRV97}). \label{l:204} If $A$ is a self-adjoint
operator,  and $A$ is affiliated with some abelian von Neuman agebra $\mathcal{R}$, then $\{E^A_\lambda\}\subseteq \mathcal{R}$.

\vskip0.1in

{\bf Lemma 9} (\cite{KRV97}). \label{l:203} If
$\{E_\lambda\}$ is a resolution of the identity, $\mathcal{R}$ is an abelian von Neumann algebra
containing $\{E_\lambda\}$, then there is a self-adjoint operator $A$ affiliated
with $\mathcal{R}$, and $$Ax=\int_{-n}^n \lambda dE_\lambda x$$
for each $x\in F_n(\mathcal{H})$ and $n\in \mathbf{N}$, where
$F_n=E_n-E_{-n}$; and $\{E_\lambda\}$ is the resolution of the
identity for $A$.

\vskip0.1in

{\bf Example} (\cite{KRV97}). \label{l:201} If
$(S,\varphi,m)$ is a $\sigma$-finite measure space and $\mathcal{A}$
is its multiplication algebra acting on $L^2(S,\varphi,m)$, then $A$ is a
closed densely defined operator affiliated with $\mathcal{A}$ if and only if $A=M_g$ (multiplication by $g$) for some measurable function $g$ finite almost
everywhere on $S$. In this case, $A$ is self-adjoint if and only if
$g$ is real-valued almost everywhere.

\vskip0.1in

Kadison and Liu (\cite{KRV12}) showed that the Heisenberg relation
$QP-PQ=-i\hbar I$ cannot be satisfied with self-adjoint operators affiliated with any finite von
Neumann algebra.

\vskip0.1in

{\bf Theorem 1.} Let $A,B\in \mathcal{S}(\mathcal{H})$. If $A\perp B$, then there exists an abelian von Neumann algebra $\mathcal{R}$ such
that $A\eta \mathcal{R}$ and $B\eta \mathcal{R}$. Moreover,
$\bigcup_{n=1}^\infty F^A_nF^B_n(\mathcal{H})$ is a common core
for $A$ and $B$.

{\bf Proof.} Suppose that $A\perp B$, that is, $\overline{ran}(A)$
is orthogonal to $\overline{ran}(B)$. It follows that $AF^A_n BF^B_m=BF^B_mAF^A_n=0$
for $m,n\in \mathbf{N}$. For each $x\in {\mathcal{D}}(B)$, as
$BF^B_mx \rightarrow Bx$, we have $AF^A_n B\subseteq BAF^A_n$. From
Lemma 4, $F^B_mAF^A_n=AF^A_nF^B_m$ for $m,n\in \mathbf{N}$.
Similarly, we have $F^A_n BF^B_m=BF^B_m F^A_n$. Also, we note that
$F^A_nF^B_m=F^B_mF^A_n$ (see Lemma 18 and Proposition 32 of \cite{Z}). Moreover, $\bigcup_{n=1}^\infty
F^A_n F^B_n(\mathcal{H})$ is a common core for $A$ and $B$.

Let $\mathcal{R}$ be the von Neumann algebra generated by
$\{F^A_n, AF^A_n, F^B_n, BF^B_n:n=1,2,\cdots\}$. Since the elements
in $\{F^A_n, AF^A_n, F^B_n, BF^B_n:n=1,2,\cdots\}$ are commuting,
$\mathcal{R}$ is abelian. If $U$ is a unitary operator in
$\mathcal{R}'$ and $x\in \bigcup_{n=1}^\infty F^A_n(\mathcal{H})$ (a core of $A$), then
$AUx=AUF^A_nx=AF^A_nUx=UAF^A_nx=UAx$ for some $n$. So $A\eta
\mathcal{R}$. Similarly, $B\eta \mathcal{R}$. \qed

\section{Generalized Orthoalgebra $({\cal
S}({\mathcal{H}}), \bot, \oplus, 0)$}

In this section, we show that $({\cal S}({\mathcal{H}}), \bot, \oplus, 0)$ is
a generalized orthoalgebra.

\vskip0.1in

{\bf Proposition 1.}\label{p:301} Let $A,B\in \mathcal{S}(\mathcal{H})$ with
$A^2=BA$. Then

(i) $\bigcup_{n=1}^\infty F_n^B(\mathcal{H})$ is a common core for
$A$ and $B$.

(ii) ${\mathcal{D}}(B)\subseteq {\mathcal{D}}(A)$.

{\bf Proof.} (i) Since $A^2$ is self-adjoint and $A^2=BA$, $BA$ is
self-adjoint and $AB=A^*B^*\subseteq (BA)^*=BA$ (Lemma 5).
Now, with $A^2=BA$ and $AB\subseteq BA$, we have
$(AF^A_m)^2=BAF^A_m\supseteq AF^A_m B$ for each $m\in \mathbf{N}$.
From Lemma 4, $F^B_n AF^A_m=AF^A_mF^B_n$. For each $x\in
{\mathcal{D}}(A)$ and $n\in \mathbf{N}$, since $AF^A_mx\rightarrow
Ax$ as $m\rightarrow\infty$, $F^B_nAx=F^B_n(\lim_m
AF^A_mx)=\lim_mF^B_nAF^A_mx=\lim_m AF^A_mF^B_nx$. Since $A$ is closed
and $F^A_mF^B_nx\rightarrow F^B_nx$ as $m\rightarrow\infty$, we
have $F^B_nx\in {\mathcal{D}}(A)$ and $AF^B_nx =F^B_nAx$. So
$F^B_nA\subseteq AF^B_n$ for each $n\in \mathbf{N}$. It follows that
$\bigcup_{n=1}^\infty F^B_n(\mathcal{H})$ is a core for $A$, hence
a common
core for $A$ and $B$.

(ii) Since $F^B_nA\subseteq AF^B_n$ (note that $AF^B_n$ is closed
since $A$ is closed and $F^B_n$ is bounded), $F^B_nA$ is closable. From Lemma 18 of \cite{Z1},
$\overline{F^B_nA}=AF^B_n$. Thus
$(F^B_nA)^*=(\overline{F^B_nA})^*=(AF^B_n)^*$ (recall that if $T$
is closable, then $\overline{T}^*=T^*$) and it follows from
Lemma 5 and Lemma 6 that
$F^B_nA\subseteq (AF^B_n)^*=(F^B_nA)^*=AF^B_n$. So $AF^B_n$ is
self-adjoint for each $n\in \mathbf{N}$. Since $A^2=BA$ and
$(AF^B_nF^B_m)^2=BF^B_nAF^B_m=AF^B_mBF^B_n$ for $m,n\in
\mathbf{N}$, we have $BF^B_nA\subseteq ABF^B_n$. By Lemma 4,
$P_ABF^B_n=BF^B_n P_A$. For each $x\in {\mathcal{D}}(B)$,
$BF^B_nx\rightarrow Bx$, $BF^B_nP_Ax=P_ABF^B_nx\rightarrow P_ABx$.
Since $F^B_nP_Ax\rightarrow P_Ax$ and $B$ is closed, we have
$P_Ax\in {\mathcal{D}}(B)$ and $BP_A x=P_ABx$. That is
$P_AB\subseteq BP_A$.

Since $AB\subseteq BA=A^2$, for each $x\in {\mathcal{D}}(B)$ with $Bx=0$, it follows that $x\in {\mathcal{D}}(A^2)$ and $A^2x=0$. Since $null(A^2)=null(A^*A)=null(A)$ (Lemma 2), we have $Ax=0$ and $null(B)\subseteq null(A)$. Then
$${\mathcal{H}}=\overline{ran}(A)\oplus(null(A)\cap \overline{ran}(B))\oplus null(B).$$
For each $x\in {\mathcal{D}}(B)$, $x=x_1+x_2+x_3$ where $x_1\in \overline{ran}(A)$, $x_2\in (null(A)\cap \overline{ran}(B))$ and $x_3\in null(B)$. Then $P_AB(x_1+x_2)=BP_A(x_1+x_2)=BP_Ax_1=Bx_1$. Thus $x_1\in {\mathcal{D}}(B)$. Since $x_1\in \overline{ran}(A)$, there exists a sequence $\{y_m\}\subseteq {\mathcal{D}}(A)$ such that $Ay_m\rightarrow x_1$. For each $n\in \mathbf{N}$, $AF^B_nx_1=AF^B_n(\lim_m Ay_m)=\lim_m AF^B_nAy_m=\lim_m BA F^B_ny_m=\lim_m BF^B_nA F^B_ny_m=BF^B_n(\lim_m F^B_nAy_m)=BF^B_nx_1\rightarrow Bx_1$. Since $A$ is closed, we have $x_1\in {\mathcal{D}}(A)$ and $Ax_1=Bx_1$. We have $x_i\in {\mathcal{D}}(A)$ for $i=1,2,3$ and hence $x\in \mathcal{D}(A)$. Therefore, ${\mathcal{D}}(B)\subseteq {\mathcal{D}}(A)$. \qed

\vskip0.1in

{\bf Proposition 2.}\label{t:301}
Let $A,B\in \mathcal{S}(\mathcal{H})$ with $A\perp B$. Then $A+B$ is densely defined and self-adjoint. (cf. Lemma 5)

{\bf Proof.}
From Theorem 1, $\bigcup_{n=1}^\infty F^A_nF^B_n(\mathcal{H})$ is a common core for $A$ and $B$. So ${\mathcal{D}}(A+B)={\mathcal{D}}(A)\bigcap {\mathcal{D}}(B)$ is dense.

To see that $A+B$ is closed, let $\{x_n\}\subseteq {\mathcal{D}}(A+B)$ with $x_n\rightarrow x$ and $(A+B)x_n\rightarrow y$. Since ${\mathcal{H}}=\overline{ran}(A)\oplus null(A)$, we have $x_n=x^{(1)}_n+x^{(2)}_n$ where $\{x^{(1)}_n\}\subseteq \overline{ran}(A)$ and $\{x^{(2)}_n\}\subseteq null(A)$. Since $\overline{ran}(A)\subseteq null(B)$ (Lemma 7), we have $(A+B)x_n= Ax^{(1)}_n+Bx^{(2)}_n\rightarrow y$.
Now $N_B(A+B)x_n=N_BAx^{(1)}_n+N_BBx^{(2)}_n=Ax^{(1)}_n+0\rightarrow N_By$, then $Bx^{(2)}_n=y-Ax^{(1)}_n\rightarrow y-N_By$. Since $A$ is closed and $Ax_n=Ax^{(1)}_n\rightarrow N_By$, it follows that $x\in {\mathcal{D}}(A)$ and $Ax=N_By$. Similarly,
since $B$ is closed and $Bx_n=Bx^{(2)}_n\rightarrow y-N_By$,
we have $x\in {\mathcal{D}}(B)$ and $Bx=y-N_By$. Therefore, $x\in {\mathcal{D}}(A)\bigcap {\mathcal{D}}(B)$ and $(A+B)x=y$, which implies that $A+B$ is closed.

We note that $\bigcup_{n=1}^\infty F^A_n F^B_n(\mathcal{H})$ is also a core for $(A+B)^*$.
To see this, since $F^A_n F^B_n(A+B)\subseteq (A+B)F^A_n F^B_n$ and $F^A_nF^B_n=F^B_nF^A_n$,
we have $F^A_n F^B_n(A+B)^*\subseteq (A+B)^*F^A_n F^B_n$.
For each $x\in {\mathcal{D}}((A+B)^*)$, $F^A_nF^B_nx\rightarrow x$ and $(A+B)^*F^A_nF^B_nx=F^A_nF^B_n(A+B)^*x\rightarrow (A+B)^*x$. So $\bigcup_{n=1}^\infty F^A_n F^B_n(\mathcal{H})$ is a core for $(A+B)^*$.

Since $A+B=A^*+B^*\subseteq (A+B)^*$ and they have the same common core, we have $A+B=(A+B)^*$ and $A+B$ is self-adjoint.\qed

\vskip0.1in
Now, for $A,B\in \mathcal{S}(\mathcal{H})$, we define $A\oplus B=A+B$ when $A\perp B$.
\vskip0.1in

{\bf Theorem 2.}\label{t:302} $({\cal S}({\mathcal{H}}), \bot,
\oplus, 0)$ is a generalized orthoalgebra.

{\bf Proof.} Clearly, the conditions (OA1) and
(GOA2) hold in $({\cal S}({\mathcal{H}}), \bot,
\oplus, 0)$.

(OA2): Let $A\perp B$ and $C\perp (A\oplus B)$. We first show that
$C\perp B$. For each $x\in {\mathcal{D}}(C)$, since
$(A+B)\perp C$, we have $(A+B)Cx=ACx+BCx=0$. Then $\langle
ACx, BCx\rangle+\langle BCx,BCx\rangle=0$. Since $A\perp B$, $\langle
ACx, BCx\rangle=0$. So $\langle BCx,BCx\rangle=0$ and $BCx=0$. Thus
$ran(C)\subseteq null(B)$. Since $null(B)$ is closed, we have
$\overline{ran}(C)\subseteq null(B)$. It follows from Lemma 7 that $C\perp B$.
Similarly, we have $C\perp A$. Next, we show that $(B+C)\perp A$. For
each $x\in {\mathcal{D}}(B+C)={\mathcal{D}}(B)\bigcap
{\mathcal{D}}(C)$, the fact that $A\perp B$ and $A\perp C$ implies that $Bx\in null(A)$
and $Cx\in null(A)$. Thus $(B+C)x\in null(A)$. Then
$\overline{ran}(B+C)\subseteq null(A)$. By  Lemma 7 again, we
obtain $(B+C)\perp A$, that is, $(B\oplus C)\perp A$. It is obvious that
$(A\oplus B)\oplus C=A\oplus (B\oplus C)$.

(OA4). Let $A\perp A$. Then $\overline{ran}(A)$ is orthogonal to $\overline{ran}(A)$. For each $x\in {\mathcal{D}}(A)$, $\langle Ax,Ax\rangle=0$ and $Ax=0$. Since ${\mathcal{D}}(A)$ is dense in $\mathcal{H}$, $A=0$.

(GOA1). Let $A\oplus B=A\oplus C$. We have proved that $\bigcup
F_n^AF_n^B(\mathcal{H})$, $\bigcup F_n^AF_n^C(H)$ are the common cores for
$A,B$ and $A,C$ respectively. Obviously, $\{F_n^AF_n^BF_n^C\}$
has strong operator limit $I$, the identity operator, and
$\bigcup_{n=1}^\infty F_n^AF_n^BF_n^C (\mathcal{H})$ is dense in $\mathcal{H}$. It
follows that $\bigcup_{n=1}^\infty F_n^AF_n^BF_n^C (\mathcal{H})$ is a common core for
$A,B$, and $C$. Since $A\oplus B=A\oplus C$, we have $(A+B)|_{\cup
F_n^AF_n^BF_n^C (\mathcal{H})}=(A+C)|_{\cup F_n^AF_n^BF_n^C (\mathcal{H})}$. Thus
$B|_{\cup F_n^AF_n^BF_n^C (\mathcal{H})}=C|_{\cup F_n^AF_n^BF_n^C (\mathcal{H})}$
and $B=C$ (since $B=C$ on their common core).

(GOA3). Suppose $A\perp B$ and $A\oplus B=0$. Then
$\overline{ran}(A)$ is orthogonal to $\overline{ran}(B)$ and $A+B=0$. For each $x\in
{\mathcal{D}}(A)\cap {\mathcal{D}}(B)$, $Ax+Bx=0$, $\langle
Ax+Bx,Ax\rangle=0$. Since $\langle Ax,Bx\rangle=0$, $\langle
Ax,Ax\rangle=0$ and $Ax=0$. Then $Ax=0$ on ${\mathcal{D}}(A)\cap {\mathcal{D}}(B)$
(dense in $\mathcal{H}$). Thus, $A=0$.
Similarly, We have $B=0$.

Therefore, $({\cal S}({\mathcal{H}}), \bot, \oplus, 0)$ is a generalized
orhtoalgebra. \qed

\section{The order properties of $({\cal S}({\mathcal{H}}), \bot,
\oplus, 0)$}

In this section, we study order properties of $({\cal S}({\mathcal{H}}), \bot,
\oplus, 0)$. For $A,B\in \mathcal{S}(\mathcal{H})$, we define $A\preceq B$
if there exists a $C\in \mathcal{S}(\mathcal{H})$ such that $A\perp C$ and
$A\oplus C=B$. It is clear $0\preceq A$ for each $A\in \mathcal{S}(\mathcal{H})$.

\vskip0.1in

{\bf Proposition 3.}\label{t:303}
Let $A,B\in \mathcal{S}(\mathcal{H})$. Then $A\preceq B$ if and only if $A^2=BA$.

{\bf Proof.}
Suppose $A\preceq B$. There is a $C\in \mathcal{S}(\mathcal{H})$ such that $A\perp C$ and $A\oplus C=B$, namely, $A+C=B$.
Let $x\in {\mathcal{D}}(A^2)$, which implies $x\in {\mathcal{D}}(A)$ and $Ax\in {\mathcal{D}}(A)$.
Since $A\perp C$, $\overline{ran}(A)\subseteq null(C)$. So $Ax\in {\mathcal{D}}(C)$.
Then $Ax\in {\mathcal{D}}(A)\bigcap {\mathcal{D}}(C)={\mathcal{D}}(B)$ and $BAx=A^2x+CAx$.
Since $CAx=0$, $BAx=A^2x$ for each $x\in {\mathcal{D}}(A^2)$.
Thus $A^2\subseteq BA$. Now, suppose $x\in {\mathcal{D}}(BA)$. Then $x\in {\mathcal{D}}(A)$ and $Ax\in {\mathcal{D}}(B)$. Since $A\perp C$, $\bigcup_{n=1}^\infty F^{A}_nF^{C}_n(\mathcal{H})$ is a common core for $A$ and $C$ and so it is also a core for $B(=A+C)$. Since $BF^{A}_nF^{C}_nAx=AF^{A}_nF^{C}_nAx+CF^{A}_nF^{C}_nAx$ and $BF^{A}_nF^{C}_nAx\rightarrow BAx$, $CF^{A}_nF^{C}_nAx\rightarrow CAx=0$, we have $AF^{A}_nF^{C}_nAx\rightarrow BAx$.
Since $F^{A}_nF^{C}_nAx\rightarrow Ax$ and $A$ is closed, we obtain $Ax\in{\mathcal{D}}(A)$ and $A^2x=BAx$. It follows that $BA\subseteq A^2$. Hence $A^2=BA$.

 Conversely, suppose $A^2=BA$. By Proposition 1, ${\mathcal{D}}(B)\subseteq {\mathcal{D}}(A)$ and ${\mathcal{D}}(B-A)$ is sense in $\mathcal{H}$. Since $B-A=B^*-A^*\subseteq(B-A)^*$, $B-A$ is symmetric which implies that $B-A$ is closable.
Define $C=\overline{B-A}$. From Proposition 1, $\bigcup_{n=1}^\infty F^B_n(\mathcal{H})$ is a common core for $A$ and $B$. So $\bigcup_{n=1}^\infty F^B_n(\mathcal{H})$ is a core for $C$, and $F^B_n C\subseteq CF^B_n$.
Then $F^B_n C^* \subseteq C^*F^B_n$, and $\bigcup_{n=1}^\infty F^B_n(\mathcal{H})$ is a core for $C^*=(\overline{B-A})^*$. Since $C=\overline{B-A}\subseteq (\overline{B-A})^*=C^*$,
we have $C=C^*$ ($Cx=C^*x$ for each $x$ in the common core
$\bigcup_{n=1}^\infty F^B_n(\mathcal{H})$ for $C$ and $C^*$) and $C$ is self-adjoint.
For $x\in {\mathcal{D}}(A)$, $F^B_nx\rightarrow x$, $AF^B_nx=F^B_nAx\rightarrow Ax$, and $A^2F^B_nx=BAF^B_nx$.
It follows that $CAF^B_nx=BAF^B_nx-A^2F^B_nx=0\rightarrow 0$. Since $C$ is closed, $Ax\in {\mathcal{D}}(C)$ and $CAx=0$. Thus $ran(A)\subseteq null(C)$ and  $\overline{ran}(A)\subseteq null(C)$. Then $C\perp A$ and $A\oplus C=A+C$. Since $B\subseteq A+(\overline{B-A})=A+C$ and $A+C$ is self-adjoint (Proposition 2),
from the fact that a self-adjoint operator is maximal symmetric, we have $B=A+(\overline{B-A})=A+C$. By definition, $A\preceq B$. \qed

\vskip0.1in
Note that from Proposition 1 and Proposition 3, $A\preceq B$ implies ${\mathcal{D}}(B)\subseteq {\mathcal{D}}(A)$.

\vskip0.1in

{\bf Proposition 4.}
 For each $A\in \mathcal{S}(\mathcal{H})$, if  $B,C\in \mathcal{S}(\mathcal{H})$ with $B,C\preceq A$ and $B\perp C$, then $B\oplus C\preceq A$. In this case, we say that $A$ is {\it principal}.

{\bf Proof.} Suppose $B,C\in \mathcal{S}(\mathcal{H})$ with $B,C\preceq A$ and $B\perp C$. It follows from Proposition 3 that $B^2=AB$ and $C^2=AC$. From Proposition 1, $\bigcup_{n=1}^\infty F^A_n(\mathcal{H})$ is a common core for $A,B$ and $C$. and therefore $A-(B+C)$ is densely defined. Since $A-(B+C)\subseteq (A-(B+C))^*$, $A-(B+C)$ is closable. Define $H=\overline{A-(B+C)}$. Just as in the proof of Proposition 3, one can prove that $H$ is self-adjoint.
For each $x\in {\mathcal{D}}(B+C)$, $F^A_nx\rightarrow x$ and $(B+C)F^A_nx=F^A_n(B+C)x\rightarrow (B+C)x$.
Since $B\perp C$, it follows that $H(B+C)F^A_nx=(A-(B+C))(B+C)F^A_nx=A(B+C)F^A_nx-(B^2+C^2)F^A_nx=0$. Since $H$ is closed, $(B+C)x\in {\mathcal{D}}(H)$ and $H(B+C)x=0$.
From Lemma 7, $H\perp (B+C)$ and $H\oplus(B+ C)=\overline{A-(B+C)}+(B+C)=A$. Hence, $B\oplus C\preceq A$ and $A$ is principle. \qed

\vskip0.1in

Recall the canonical ordering on $\mathcal{S}(\mathcal{H})$. We say that $A\leq B$ if ${\mathcal{D}}(B)\subseteq {\mathcal{D}}(A)$ and $\langle Ax,x\rangle\leq \langle Bx,x\rangle$ for each $x\in{\mathcal{D}}(B)$. Regarding
$\leq$ and the newly defined ordering $\preceq$, we have the following results:

\vskip0.1in

{\bf Proposition 5.}
If $A\preceq B$ and $B\geq 0$, then $A\leq B$.

{\bf Proof.}
Suppose $A\preceq B$. Then ${\mathcal{D}}(B)\subseteq {\mathcal{D}}(A)$ and there exists a $C\in \mathcal{S}(\mathcal{H})$ such that $A\perp C$ and $A\oplus C=B$.
For $x\in{\mathcal{D}}(B)$, $x=y+z$ where $y\in \overline{ran}(A)$ and $z\in null(A)$. Then $x,z
\in {\mathcal{D}}(A)$ implies $y\in {\mathcal{D}}(A)$, and
$\overline{ran}(A)\subseteq null(C)$ implies $y\in {\mathcal{D}}(C)$. Hence $y\in {\mathcal{D}}(A)\cap {\mathcal{D}}(C)$. If follows that $y\in {\mathcal{D}}(B)$ and $z\in {\mathcal{D}}(B)$. Then $\langle (B-A)x,x\rangle=\langle (B-A)(y+z),y+z\rangle=\langle (B-A)z,y+z\rangle=\langle z,(B-A)(y+z)\rangle=\langle z,Bz\rangle\geq 0$. So $\langle Ax,x\rangle\leq \langle Bx,x\rangle$ for each $x\in {\mathcal{D}}(B)$. Hence, $A\leq B$. \qed

\vskip0.1in

Let $\mathcal{B}(\mathbf{R})$ be the set of all Borel subsets of $\mathbf{R}$.
We now characterize the ordering $\preceq$ on $\mathcal{S}(\mathcal{H})$ in terms of the spectral measure of self-adjoint operators.
\vskip0.1in

{\bf Lemma 10.} (\cite{GS06}) For $A, B\in \mathcal{S}_{\bf b}(\mathcal{H})$, $A\preceq B$
(that is, there exists a $C \in \mathcal{S}_{\bf b}(\mathcal{H})$ such that $A \perp C$
and $A\oplus C=B$) if and only if $P^A(\Delta)\leq P^B(\Delta)$ for every $\Delta \in \mathcal{B}(\mathbf{R})$ with $0\notin \Delta$.

\vskip0.1in

{\bf Theorem 3.}\label{t:304}
Let $A,B\in \mathcal{S}(\mathcal{H})$. Then $A\preceq B$ if and only if
$E^A_{\Delta\lambda_j}\leq E^B_{\Delta\lambda_j}$, where $ E^A_{\Delta\lambda_j}=E^A_{\lambda_j}-E^A_{\lambda_{j-1}}$, $0\notin(\lambda_{j-1},\lambda_j]$, $j=1,2,3\cdots$, and $\{E^A_\lambda\}$ is the resolution of the identity for $A$.

{\bf Proof.}
Suppose $A\preceq B$. Then there exists a $C\in \mathcal{S}(\mathcal{H})$
such that $A\perp C$ and $A\oplus C=B$, and $\bigcup_{n=1}^\infty F^A_nF^C_n(\mathcal{H})$
is a common core for $A$ and $C$ (Theorem 1), and therefore a core for $B$.
From Proposition 3, we have $A^2=BA$. Then $(AF^A_nF^C_n)^2=(BF^A_nF^C_n)(AF^A_nF^C_n)$
for each $n\in \mathbf{N}$. Again, by Proposition 3, $AF^A_nF^C_n\preceq BF^A_nF^C_n$
for each $n\in \mathbf{N}$. Since $\{E^A_\lambda F^A_nF^C_n\}$ and $\{E^B_\lambda F^A_nF^C_n\}$ are the resolutions of the identity for $AF^A_nF^C_n|_{F^A_nF^C_n(\mathcal{H})}$ and $BF^A_nF^C_n|_{F^A_nF^C_n(\mathcal{H})}$,
respectively. By Lemma 10, we have $ E^A_{\Delta \lambda_j}F^A_nF^C_n\leq E^B_{\Delta  \lambda_j}F^A_nF^C_n$ for each $n\in \mathbf{N}$ and $0\notin (\lambda_{j-1},\lambda_j]$.
Since $\{F^A_nF^C_n\}$ has strong operator limit $I$, it follows that $E^A_{\Delta \lambda_j}\leq E^B_{\Delta \lambda_j}$, $0\notin (\lambda_{j-1},\lambda_j]$.

Suppose that $E^A_{\Delta \lambda_j}\leq E^B_{\Delta \lambda_j}$, $0\notin(\lambda_{j-1},\lambda_j]$.
For each $n\in \mathbf{N}$, $\{E^A_\lambda F^A_n\}$ and $\{E^B_\lambda F^B_n\}$
are the resolutions of the identity for $AF^A_n|_{F^A_n(\mathcal{H})}$ and $BF^B_n|_{F^B_n(\mathcal{H})}$, respectively. For a fixed $n\in \mathbf{N}$, for each $(\lambda_{j-1},\lambda_j]$
not containing 0, either $E^A_{\Delta \lambda_j}F^A_n=0$ and $E^B_{\Delta \lambda_j}F^B_n=0$,
or $E^A_{\Delta \lambda_j}F^A_n\leq E^B_{\Delta \lambda_j}F^B_n$. Then $AF^A_n \preceq BF^B_n$ and $(AF^A_n)^2=(BF^B_n)(AF^A_n)$. For each $x\in {\mathcal{D}}(A^2)$, $x\in{\mathcal{D}}(A)$ and $Ax\in {\mathcal{D}}(A)$. As $F^B_nF^A_n Ax\rightarrow Ax$, we have $BF^B_nF^A_nAx=BF^B_nAF^A_nx=(AF^A_n)^2x=F^A_nA^2x\rightarrow A^2x$. Since $B$ is closable, $Ax\in {\mathcal{D}}(B)$ and $BAx=A^2x$. So $A^2\subseteq BA$. Conversely, for each $x\in {\mathcal{D}}(BA)$, $x\in{\mathcal{D}}(A)$ and $Ax\in {\mathcal{D}}(B)$.  As $(AF^A_n)^2$ is self-adjoint, we have $(BF^B_n)(AF^A_n)=(AF^A_n)(BF^B_n)$. By Lemma 4, we have $BF^B_nF^A_n=F^A_n BF^B_n$. Since $F^A_nAx\rightarrow Ax$, we have $AF^A_nAx=(AF^A_n)^2x=(BF^B_n)(AF^A_n)x=F^B_nBF^A_nAx=F^B_nF^A_nBAx\rightarrow BAx$. As $A$ is closable, $Ax\in {\mathcal{D}}(A)$ and $A^2x=BAx$. So $BA\subseteq A^2$.
Therefore, $A^2=BA$ which implies $A\preceq B$. \qed

\vskip0.1in

{\bf Corollary 1.}\label{c:301}
Let $A,B\in \mathcal{S}(\mathcal{H})$. Then $A\preceq B$ if and only if $P^A(\Delta)\leq P^B(\Delta)$ for every $\Delta \in \mathcal{B}(\mathbf{R})$ with $0\notin \Delta$.

\vskip0.1in

Next, we study the existence of $A\wedge B$ and $A\vee B$ for $A,B\in \mathcal{S}(\mathcal{H})$.
For each $\Delta\in \mathcal{B}(\mathbf{R})$, if $\Delta=\cup_{i=1}^n\Delta_i$, where $\{\Delta_i\}_{i=1}^n$ are pairwise disjoint sets in $\mathcal{B}(\mathbf{R})$, then we say $\gamma=\{\Delta_i\}_{i=1}^n$ is a partition of $\Delta$. We denote all the partitions of $\Delta$ by $\Gamma(\Delta)$.
Let $A, B\in \mathcal{S}(\mathcal{H})$. Define $P:{\mathcal{B}(\mathbf{R})}\rightarrow P(\mathcal{H})$
as follows. Let $P(\emptyset)=0$, and for each nonempty $\Delta\in {\mathcal{B}}(\mathbf{R})$ and $\gamma\in \Gamma(\Delta)$,

$$
P(\Delta)=\left\{ \begin{array}{ccc}
    \wedge_{\gamma\in \Gamma(\Delta)}\sum_{\Delta_i\in \gamma}(P^A(\Delta_i)\wedge P^B(\Delta_i)), &\ \ & 0\not\in\Delta  \\
    I-P({\mathbf{R}}\setminus\Delta). &\ \ & 0\in\Delta \\
\end{array}\right..$$
\vskip0.1in

{\bf Lemma 11} (\cite{SJZ09}). As defined above, $P:{\mathcal{B}(\mathbf{R})}\rightarrow {\cal P}(\mathcal{H})$ is a spectral measure.

\vskip0.1in

{\bf Theorem 4.}
Let $A, B\in \mathcal{S}(\mathcal{H})$. Then $A\wedge B$ exists in $\mathcal{S}(\mathcal{H})$ with respect to $\preceq$.

{\bf Proof.}
Let $\{E^A_\lambda\}$, $\{E^B_\lambda\}$ be the resolutions of identity for $A$ and $B$, $P^A$ and $P^B$ be the spectral measures for $A$ and $B$,  respectively. Define $P(\Delta)$ as above for each Borel set $\Delta\in \mathcal{B}(\mathbf{R})$ and then $P$ is a spectral measure. Define $E_\lambda=P((-\infty,\lambda])$ and $\{E_\lambda\}$ is a resolution of identity. By Lemma 9, there exists a self-adjoint
operator $C$ such that $$Cx=\int_{-n}^n \lambda dE_\lambda x,$$
where $x\in F_n(\mathcal{H})$, $n\in \mathbf{N}$, $F_n=E_n-E_{-n}$, and $\{E_\lambda\}$ is the resolution of
the identity for $C$.
Let $\Delta \in\mathcal{B}(\mathbf{R})$ with $0\notin \Delta$. For each $\gamma\in \Gamma (\Delta)$,
\begin{eqnarray*}
P_\gamma &=& \sum_{\Delta_i\in \gamma}(P^A(\Delta_i)\wedge P^B(\Delta_i))\\
&\leq& (\sum_{\Delta_i\in \gamma}(P^A(\Delta_i)))\wedge (\sum_{\Delta_i\in \gamma}(P^B(\Delta_i)))\\
&=& P^A(\Delta)\wedge P^B(\Delta).
\end{eqnarray*}
Then $P(\Delta)=\wedge_{\gamma\in \Gamma(\Delta)}P_\gamma\leq P^A(\Delta)\wedge P^B(\Delta)$.
From Corollary 1, $C\preceq A$ and $C\preceq B$. Suppose there exists another $C_1\in \mathcal{S}(\mathcal{H})$ such that $C_1\preceq A$ and $C_1\preceq B$. For each $\Delta\in \mathcal{B}(\mathbf{R})$ with $0\notin \Delta$ and $\gamma\in \Gamma(\Delta)$, since $P^{C_1}(\Delta_i)\leq P^A(\Delta_i)$ and $P^{C_1}(\Delta_i)\leq P^B(\Delta_i)$ for each Borel subsets $\Delta_i\in \gamma$, we have $$P^{C_1}(\Delta)=\sum_{\Delta_i\in\gamma}P^{C_1}(\Delta_i)\leq \sum_{\Delta_i\in\gamma}P^A(\Delta_i)\wedge P^B(\Delta_i).$$ So we obtain $$P^{C_1}(\Delta)\leq\wedge_{\gamma\in \Gamma(\Delta)}\sum_{\Delta_i\in\gamma}P^A(\Delta_i)\wedge P^B(\Delta_i)=P(\Delta).$$ Therefore, $C_1\preceq C $ and $C=A\wedge B$. \qed

\vskip0.1in
{\bf Remark 1.}
 If $\{A_\alpha\}_{\alpha\in\Lambda}\subseteq \mathcal{S}(\mathcal{H})$, then $A=\wedge_\alpha A_\alpha$ exists in $\mathcal{S}(\mathcal{H})$. In fact, define $P(\emptyset)=0$  and for each nonempty $\Delta\in
\mathcal{B}({\mathbf{R}})$,
$$
P(\Delta)=\left\{ \begin{array}{ccc}
   \wedge_{\gamma\in \Gamma(\Delta)}
\sum_{\Delta_i\in \gamma}\Big(\wedge_\alpha P^{A_\alpha}(\Delta_i)\Big) &\ \ & 0\not\in\Delta  \\
    I-P({\mathbf{R}}\setminus\Delta) &\ \ & 0\in\Delta \\
\end{array}\right..$$
It can be proved that $P: {\mathcal{B}}({\mathbf{R}})\rightarrow P(\mathcal{H})$ is
a spectral measure (\cite{SJZ09}). Let $E_\lambda=P((-\infty,\lambda])$ and $\{E_\lambda\}$
is a resolution of the identity. Then we
have $A=\wedge_\alpha A_\alpha$, where $Ax=\int_{-n}^n\lambda dE_\lambda x$, for each $x\in F_n(\mathcal{H})$ and $n\in \mathbf{N}$.

\vskip0.1in

With $A,B\in \mathcal{S}(\mathcal{H})$, now we know that $A\preceq B$ implies $P^A(\Delta)\leq P^B(\Delta)$ for each $\Delta \in {\mathcal{B}}({\mathbf{R}})$ with $0\notin \Delta$. We have $P^A(\Delta)=P^A(\Delta) P^B(\Delta)=P^B(\Delta)P^A(\Delta)$ and $P^A(\Delta_1)P^A(\Delta_2)=0$ for
$\Delta_1, \Delta_2\in {\mathcal{B}}({\mathbf{R}})$ with $\Delta_1\cap\Delta_2=\emptyset$. The following result is straightforward.

\vskip0.1in
{\bf Lemma 12.}\label{l:303}
Let $A,B\in \mathcal{S}(\mathcal{H})$. Suppose that $H\in \mathcal{S}(\mathcal{H})$ is an upper bound of $A$ and $B$ with respect to $\preceq$. Then, for any $\Delta_1, \Delta_2 \in {\mathcal{B}}({\mathbf{R}})$ with $\Delta_1\cap\Delta_2=\emptyset$ and $0\notin \Delta_1\cup\Delta_2$, we have $$P^A(\Delta_1)P^B(\Delta_2)=P^A(\Delta_1)P^H(\Delta_1)P^H(\Delta_2)P^B(\Delta_2)=0.$$

\vskip0.1in

{\bf Lemma 13} (\cite{LWH09}). \label{l:304}
Let $A,B\in \mathcal{S}(\mathcal{H})$. Suppose that $P^A(\Delta_1)P^B(\Delta_2)=0$ for each pair $\Delta_1,\Delta_2\in \mathcal{B}(\mathbf{R})$ with $\Delta_1\cap\Delta_2=\emptyset$ and $0\notin \Delta_1\cup\Delta_2$. Then the following mapping $P:{\mathcal{B}(\mathbf{R})}\rightarrow P(\mathcal{H})$ defines a spectral measure:
$$
P(\Delta)=\left\{ \begin{array}{ccc}
    P^A(\Delta)\vee P^B(\Delta), &\ \ & 0\not\in\Delta  \\
    P^A(\Delta\backslash\{0\})\vee P^B(\Delta\backslash\{0\})+N_A\wedge N_B. &\ \ & 0\in\Delta \\
\end{array}\right..$$

\vskip0.1in

{\bf Theorem 5.}
Let $A,B\in \mathcal{S}(\mathcal{H})$. If there exists a $C\in \mathcal{S}(\mathcal{H})$ such that $A\preceq C$ and $B\preceq C$, then $A\vee B$ exists in $\mathcal{S}(\mathcal{H})$ with respect to $\preceq$.

{\bf Proof.}
Define $P$  as  in Lemma 13. Then $P$ is a spectral measure and $\{E_\lambda\}$ is a resolution of the identity, where $E_\lambda=P((-\infty,\lambda])$. By Lemma 9, there exists a self-adjoint operator $C$ such that $$Cx=\int_{-n}^n \lambda dE_\lambda x,$$
where $x\in F_n(\mathcal{H})$, $n\in \mathbf{N}$, $F_n=E_n-E_{-n}$,
and $\{E_\lambda\}$ is the resolution of the identity for $C$.
Clearly, $P^A(\Delta)\leq P(\Delta)$ and $P^B(\Delta)\leq P(\Delta)$ for each
$\Delta \in \mathcal{B}(\mathbf{R})$ with $0\notin \Delta$. It follows from Corollary 1, $A\preceq C$ and $B\preceq C$. If there exists another $C_1\in \mathcal{S}(\mathcal{H})$ such that $A\preceq C_1$ and $B \preceq C_1$. Then $P^A(\Delta)\leq P^{C_1}(\Delta)$, $P^B(\Delta)\leq P^{C_1}(\Delta)$ and $P^A(\Delta)\vee P^B(\Delta)\leq P^{C_1}(\Delta)$ for each $\Delta\in \mathcal{B}(\mathbf{R})$ with $0\notin \Delta$. Then $P^C(\Delta)\leq P^{C_1}(\Delta)$ for each $\Delta\in \mathcal{B}(\mathbf{R})$ with $0\notin \Delta$. Therefore, by Corollary 1, $C\preceq C_1$ and $C=A\vee B$. \qed

\vskip0.1in

{\bf Remark 2.}
Let $\{A_\alpha\}_{\alpha\in\Lambda}\subseteq \mathcal{S}(\mathcal{H})$ and $A_\alpha\preceq H$ for each $\alpha\in \Lambda$. Then $A=\bigvee_\alpha A_\alpha$ exists in $\mathcal{S}(\mathcal{H})$. In fact, define

$$
P(\Delta)=\left\{ \begin{array}{ccc}
    \vee_\alpha P^{A_\alpha}(\Delta), &\ \ & 0\not\in\Delta  \\
    \vee_\alpha P^{A_\alpha}(\Delta\backslash\{0\})+\wedge_\alpha N_{A_\alpha}.&\ \ & 0\in\Delta \\
\end{array}\right..$$
It can be proved that $P:\mathcal{B}(R)\rightarrow P(\mathcal{H})$ defines a spectral measure.
Then $\{E_\lambda\}$, where $E_\lambda=P((-\infty,\lambda])$, is a resolution of the identity.  There exists a self-adjoint operator $A$ such that $Ax=\int_{-n}^n \lambda dE_\lambda x$ for each $x\in F_n(\mathcal{H})$ and $n\in \mathbf{N}$. Then $\vee_\alpha A_\alpha=A$.

\vskip0.1in

{\bf Theorem 6.} Let ${\mathcal{H}}=L^2(-\infty,+\infty)$. Then $Q\wedge P=0$ with
respect to the order $\preceq$, where $Q$ and $P$ are the position operator $Q$ and momentum operator
$P$ satisfying the Heisenberg's commutation relation $QP-PQ=-i\hbar I$.

{\bf Proof. } Suppose that there exists an $A\in \mathcal{S}(\mathcal{H})$
such that $A\preceq P$ and $A\preceq Q$. By Proposition 3, $A^2=PA$ and $A^2=QA$.
It follows that $A^3=PA^2=PQA$ $A^3=QA^2=QPA$, and therefore $PQA=QPA$. Applying Heisenberg's commutation relation $QP-PQ=-i\hbar I$, we have $$QPA-PQA=(QP-PQ)A=-i\hbar IA.$$
Since $\bigcup_{n=1}^\infty F^A_n({\mathcal{H}})\subseteq {\mathcal{D}}(A^3)={\mathcal{D}}(PQA)={\mathcal{D}}(QPA)$,  $QPAx-PQAx=(QP-PQ)Ax=-i\hbar IAx$ for each $x\in \bigcup_{n=1}^\infty F^A_n(\mathcal{H})$. So $Ax=0$ for each $x\in \bigcup_{n=1}^\infty F^A_n(\mathcal{H})$, which implies that $A=0$. Therefore, we have $Q\wedge P=0$. \qed

\bibliographystyle{amsplain}

\end{document}